\documentclass[a4paper,11pt]{article}

\usepackage{pos,aas_macros,subfigure,amsmath,hyperref}

\title{Chasing Gravitational Waves with the Cherenkov Telescope Array}

\author*[a]{J. G. Green}
\author[b]{A. Carosi}
\author[c]{L. Nava}
\author[b,d]{B. Patricelli}
\author[e]{F. Schüssler}
\author[f]{M. Seglar-Arroyo}
\author[b]{A. Stamerra}

\onbehalf{for the CTA Consortium}

\affiliation[a]{Max-Planck-Institut für Physik,
Föhringer Ring 6, D-80805 Munich, Germany}

\affiliation[b]{INAF - Osservatorio Astronomico di Roma, 
Via Frascati 33, I-00078 Monte Porzio Catone (Rome), Italy}

\affiliation[c]{INAF - Osservatorio Astronomico di Brera, 
Via E. Bianchi 46, I-23807 Merate (LC), Italy}

\affiliation[d]{University of Pisa, 
Largo B. Pontecorvo 3, I-56127 Pisa, Italy}

\affiliation[e]{IRFU, CEA, Université Paris-Saclay, Bât 141, 
91191 Gif-sur-Yvette, France}

\affiliation[f]{IFAE, The Barcelona Institute of Science and Technology, 
Campus UAB, 08193 Bellaterra (Barcelona), Spain}

\emailAdd{jgreen@mpp.mpg.de}

\abstract{
  The detection of gravitational waves (GWs) from a binary neutron star (BNS) merger by Advanced LIGO and Advanced Virgo (GW170817), along with the discovery of the electromagnetic counterparts of this GW event, ushered in a new era of multimessenger astronomy, providing the first direct evidence that BNS mergers are progenitors of short gamma-ray bursts (GRBs). Such events may also produce very-high-energy (VHE, > 100GeV) photons which have yet to be detected in coincidence with a GW signal. The Cherenkov Telescope Array (CTA) is a next-generation VHE observatory which aims to be indispensable in this search, with an unparalleled sensitivity and ability to slew anywhere on the sky within a few tens of seconds. Achieving such a feat will require a comprehensive real-time strategy capable of coordinating searches over potentially very large regions of the sky. This work will evaluate and provide estimations on the number of GW-CTA events determined from simulated BNS systems and short GRBs, considering both on- and off-axis emission. In addition, we will present and discuss the prospects of potential follow-up strategies with CTA.
}

\ConferenceLogo{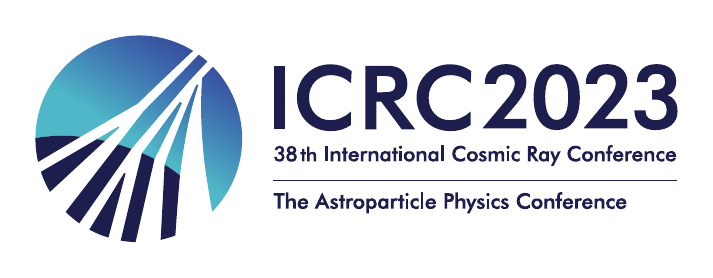}

\FullConference{%
38th International Cosmic Ray Conference (ICRC2023)\\
  26 July - 3 August, 2023\\
  Nagoya, Japan}

\begin{document}
\maketitle

\section{Introduction}

On 17 August 2017, a groundbreaking observation was made when a gravitational wave (GW) signal from the inspiral of a binary neutron star (BNS) merger was detected for the first time by Advanced LIGO and Advanced Virgo~\cite{2015CQGra..32g4001L, 2015CQGra..32b4001A}.
Approximately 2 seconds after this GW event, Fermi and INTEGRAL detected a short Gamma-Ray Burst (GRB), GRB 170817A~\cite{2017ApJ...848L..14G, 2017ApJ...848L..15S}. The association of GW170817 and GRB 170817A provided the first direct evidence that BNS mergers can produce short GRBs~ \cite{2017ApJ...848L..13A}.

Recent observations have shown that GRBs can emit very-high-energy (VHE, E $>$ 100 GeV) photons, as evidenced by the observations of GRB 190114C, GRB 160821B, GRB 201216C, and GRB 201015A by MAGIC, GRB 180720B and GRB 190829A by H.E.S.S., and GRB 221009A by LHAASO \cite{2019Natur.575..455M, 2021ApJ...908...90A, 2020GCN.29075....1B, 2020GCN.28659....1B, 2019Natur.575..464A, 2021Sci...372.1081H, 2022GCN.32677....1H}. However, no VHE emission in association with a GW event has been observed so far. Still, the hint of TeV emission from the short GRB 160821B is quite interesting — the class of GRB associated to GW emission from BNS. Detecting VHE emission associated with GW events could provide crucial insights into the physics of these astrophysical phenomena.

The next-generation ground-based observatory for gamma-ray astronomy is the Cherenkov Telescope Array (CTA). CTA is poised to play a significant role in the search for VHE emission associated with GW events as a result of its large field-of-view (FOV), quick slewing times, and unparalleled sensitivity when compared with the current generation of VHE facilities. CTA will consist of two arrays, one in each hemisphere, providing all-sky coverage. The arrays will include a mix of large (LST), medium (MST), and small size telescopes (SST), each encompassing distinct energy ranges: 20 GeV - 150 GeV, 150 GeV - 5 TeV, and 5 TeV - 300 TeV, respectively.

In this paper, we discuss the current status of the GW follow-up strategy of CTA, defined by simulating nearly the entire physical and observational process, from the BNS events themselves to their VHE observations with the most recent CTA configuration. The Advanced LIGO, Advanced Virgo, and KAGRA (LVK) network 
will operate with increased sensitivity during the fifth observing run (O5), slated to begin in 2027, allowing for a much larger volume of the Universe to be probed \cite{2020LRR....23....3A, 2022ApJ...924...54P}. This would naturally lead to an increase in the GW detection rate and likely a corresponding spike in the multi-messenger detection rate \cite{2022MNRAS.513.4159P}. Understanding the VHE emission associated with GW events can have far-reaching implications in astrophysics, helping to unravel the mysteries of the most energetic events in the Universe.

\section{Simulated Gravitational Wave Events}

In this study, we utilize a comprehensive set of simulated BNS mergers and their associated GW signals, which has been made publicly available \cite{2022ApJ...924...54P, 2021zndo...5206853S}. The simulations take into account realistic astrophysical distributions such as masses, spins, distances, and sky locations of the neutron stars. The masses of the neutron star components are drawn from a normal distribution with a mean of 1.33 M$_\odot$ and a standard deviation of 0.09 M$_\odot$, in line with measurements from binary systems in the Galaxy \cite{2016ARA&A..54..401O}. The spins of the neutron stars are either aligned or anti-aligned, with magnitudes uniformly distributed up to 0.05.

The positions and orientations of the binaries are isotropically distributed, and the redshifts are uniformly distributed in co-moving rate density, using cosmological parameters from the literature \cite{2016A&A...594A..13P}. The GW signals from the inspiral phase of these BNS mergers are simulated and convolved with the responses of different GW detector networks. For the purpose of this study, we focus on simulations that assume a LVK network, with expected sensitivities for O5 \cite{2020LRR....23....3A}.

The simulated data are analyzed using the matched filtering technique, a method optimal for extracting signals from noisy data. A 70\% independent duty cycle is assumed for each interferometer and GWs are considered to be detected if the network-wide signal-to-noise ratio is above 8, even if observed by a single interferometer. This technique reconstructs alerts characteristic of the public GW alerts sent during the third LVC observing run (O3) \cite{2022ApJ...924...54P}.

Lastly, for each simulated GW candidate, the associated sky localization is estimated with BAYESTAR, a Bayesian position reconstruction code~\cite{2016PhRvD..93b4013S}, information crucial for planning any follow-up strategy. 

\begin{figure}[ht]
  \centering
  \subfigure[]{\includegraphics[width=0.49\textwidth]{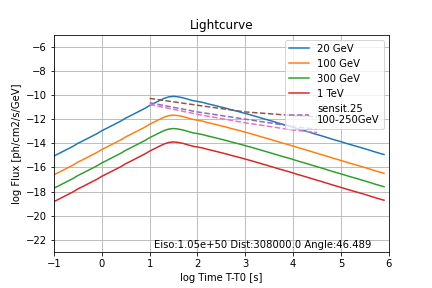}}
  \subfigure[]{\includegraphics[width=0.49\textwidth]{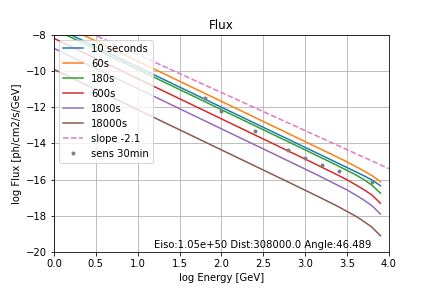}}
  \caption{Lightcurves at different energies (a) and spectra at different latencies (b) for a simulated short GRB associated to a GW event. The physical and geometrical parameters are shown in the figure: Isotropic energy in erg/s; distance in kpc; jet viewing angle in deg. For reference, the sensitivities of CTA are reported: in (a) the dashed lines show the CTA sensitivities at energies of 25, 100 and 250 GeV; in (b) the points represent the CTA differential sensitivity for an integration time of 30 minutes. 
  }
  \label{fig:LCID1378}
\end{figure}

\section{Estimation of VHE gamma-ray Emission}

It is postulated that each BNS merger in the simulated catalog results in a short GRB launching a relativistic jet. We employ a phenomenological approach that does not necessitate specifying the particle population or the exact radiative mechanism for the production of gamma-rays. The approach is based mainly on the limited information available, including detections of GRBs in the TeV range and flux upper limits from observations by Imaging Atmospheric Cherenkov Telescopes (IACTs). The luminosity in the TeV range is assumed to be comparable to that in the soft X-ray band, and the spectra are assumed to have a photon index around -2.2 without strong evidence for spectral evolution.

The isotropic equivalent energy $E_{\rm iso}$ is randomly assigned to each jet launched by a BNS merger and follows the distribution inferred from the population study presented in \cite{2016A&A...594A..84G}. This energy is defined as the isotropic equivalent energy emitted in the prompt phase as inferred by an on-axis observer. 
The jet structure is assumed to have a Gaussian distribution for both energy and Lorentz factor ($\Gamma(\theta)\propto e^{-\theta^2/2\theta_c^2}$) as a function of the angle $\theta$ from the jet axis. The 1-$\sigma$ width
$\theta_c$ of the jet (jet opening angle, hereafter) and the initial Lorentz factor are assigned based on distributions inferred from the population of short GRBs observed on-axis. 
Calculations of the evolution of the Lorentz factor require specifying the density of the external medium, which is assumed to be constant and equal to 0.1\,cm$^{-3}$.
The emission received by an observer located at an angle $\theta_{\rm view}$ from the jet axis is estimated using the methodology described in \cite{2003ApJ...592.1002S,2017MNRAS.472.4953L}. 
Figure~\ref{fig:LCID1378} shows example lightcurves and spectra simulated from a short GRB associated with a GW event.

\section{The CTA follow-up strategy}
\textbf{Estimating exposure times from simulations}

Although VHE electromagnetic (EM) emission from BNS mergers is expected to commence shortly after the merger event, there is always a delay before the EM follow-up observations can begin. This delay, known as latency, arises due to several factors, including the time required to send the GW alert to astronomers, the time needed for the telescopes to be pointed towards the region of interest, and the uncertainty in the sky location of the GW event.

\begin{figure}[ht!]
  \centering
  \subfigure[CTA South, $\theta_{\text{view}} < 10^{\circ}$]{\includegraphics[width=0.45\textwidth]{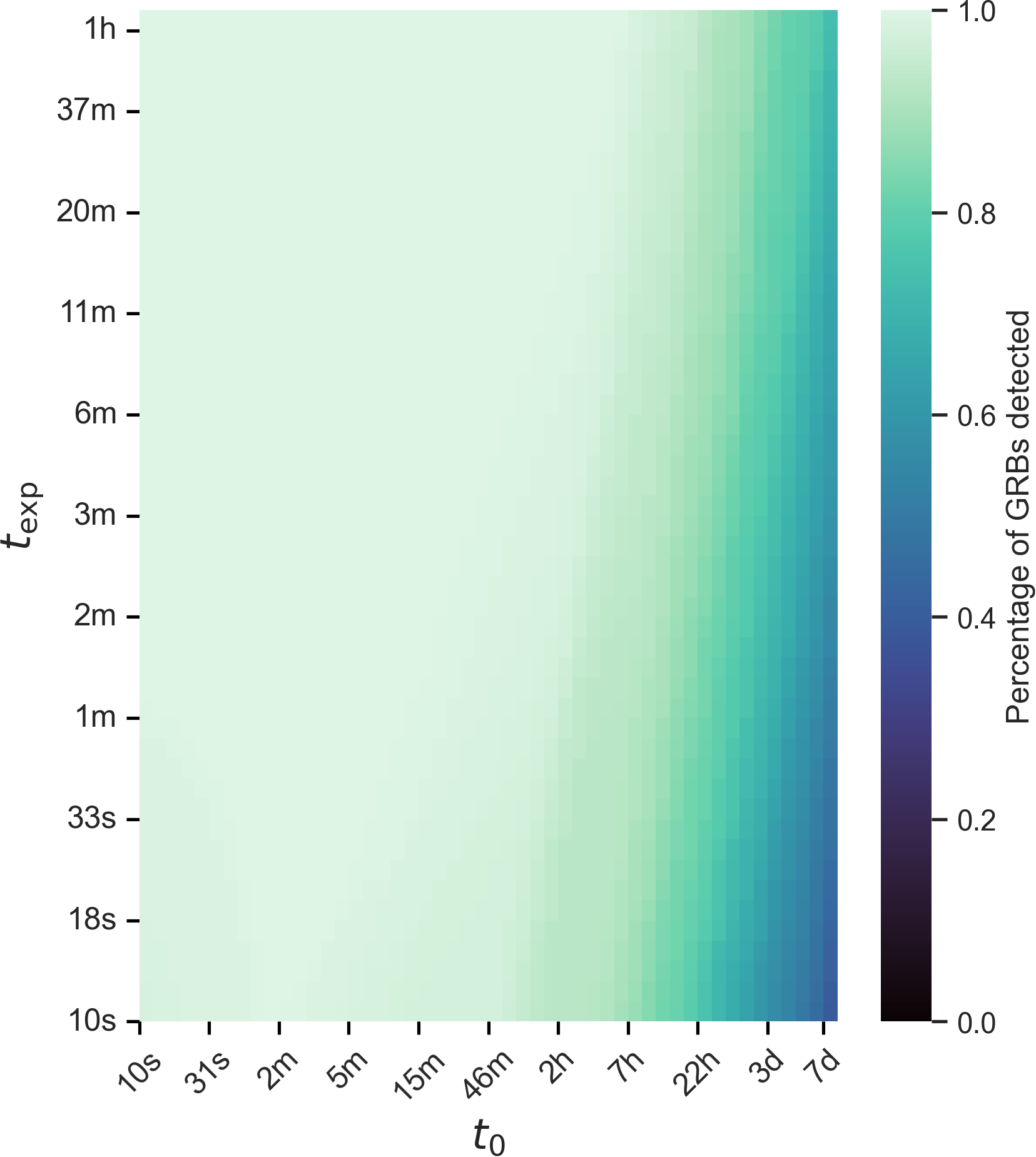}}
  \subfigure[CTA South, $\theta_{\text{view}} < 45^{\circ}$]{\includegraphics[width=0.45\textwidth]{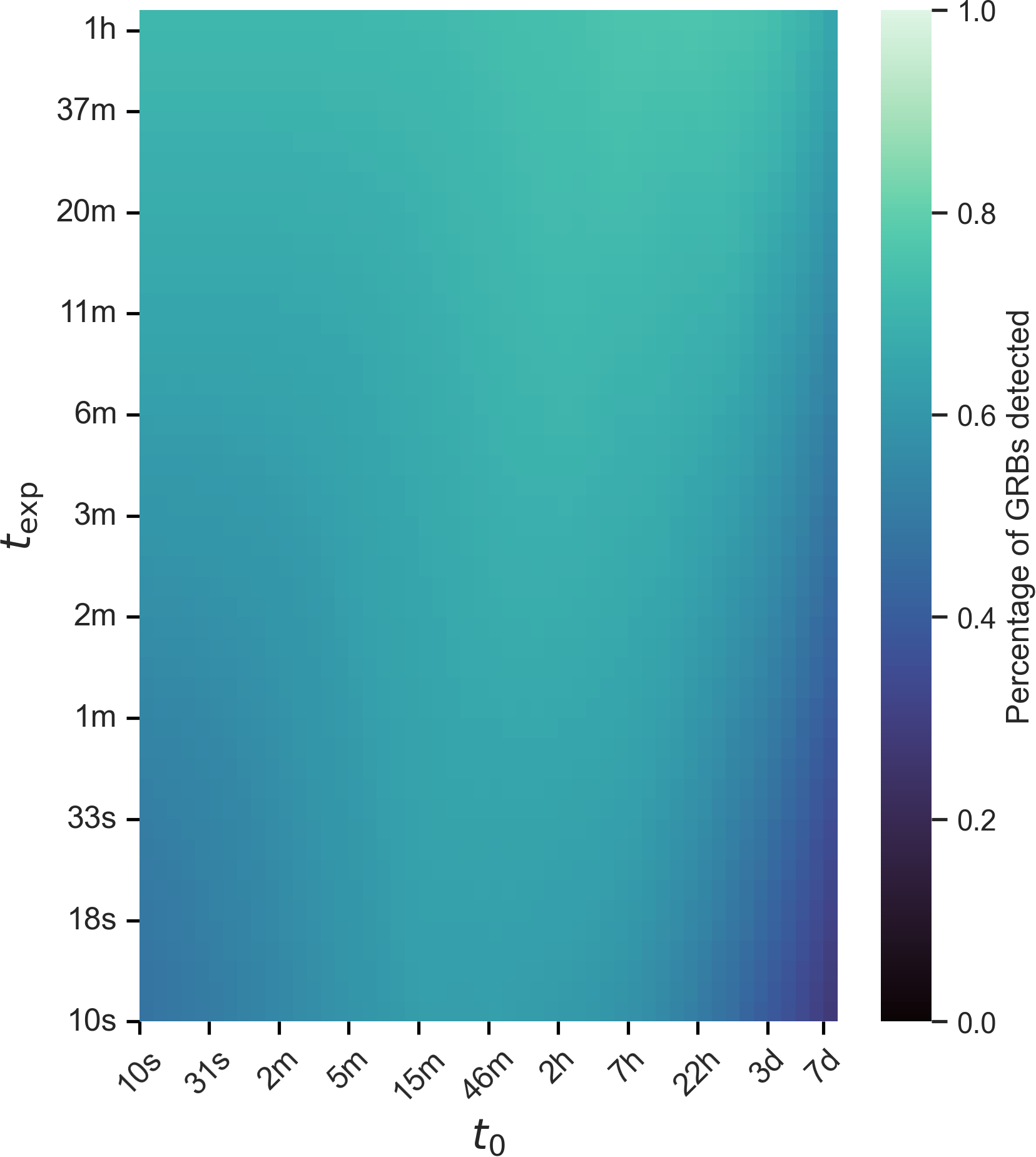}} \\
  \subfigure[CTA North, $\theta_{\text{view}} < 10^{\circ}$]{\includegraphics[width=0.45\textwidth]{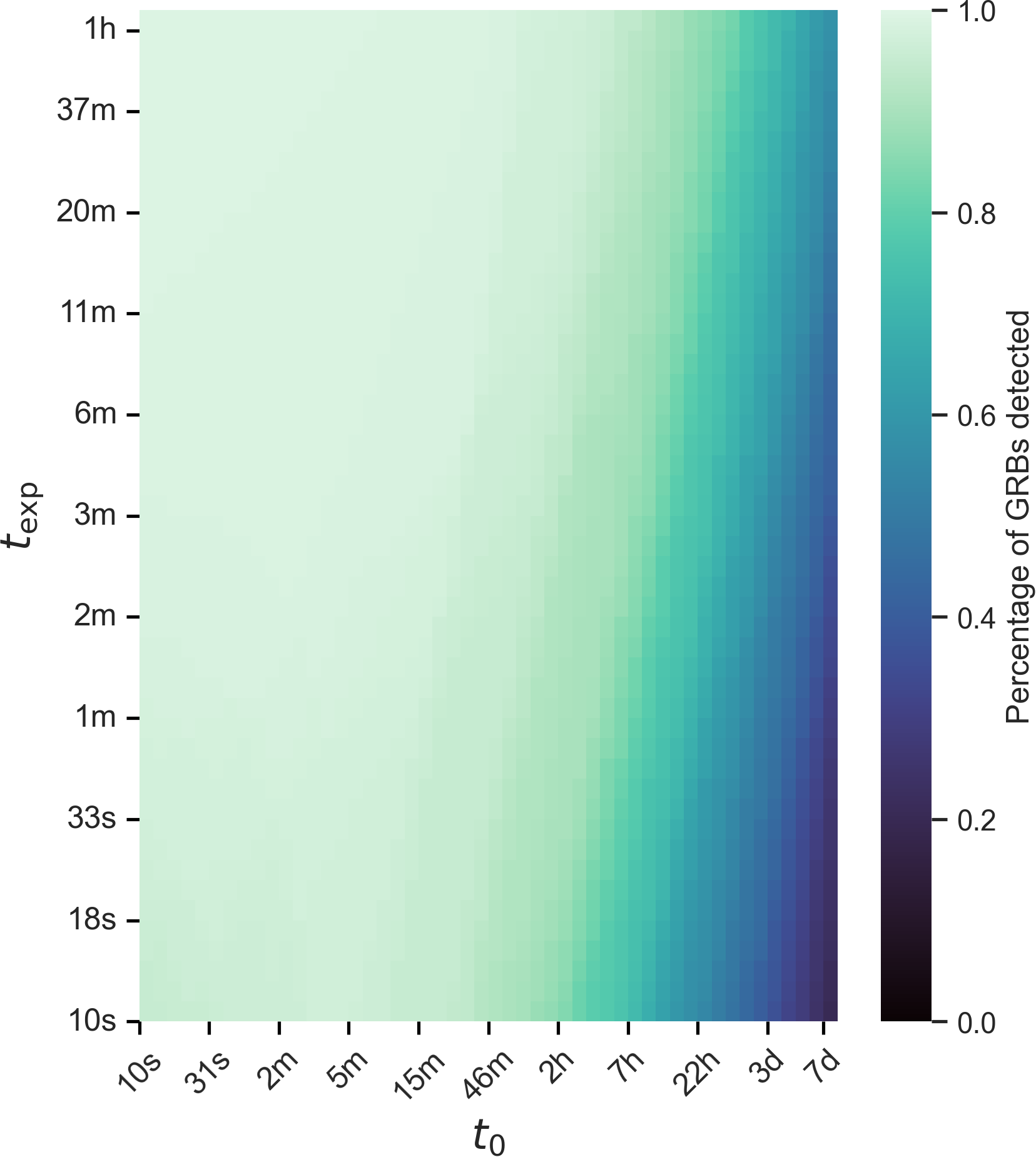}}
  \subfigure[CTA North, $\theta_{\text{view}} < 45^{\circ}$]{\includegraphics[width=0.45\textwidth]{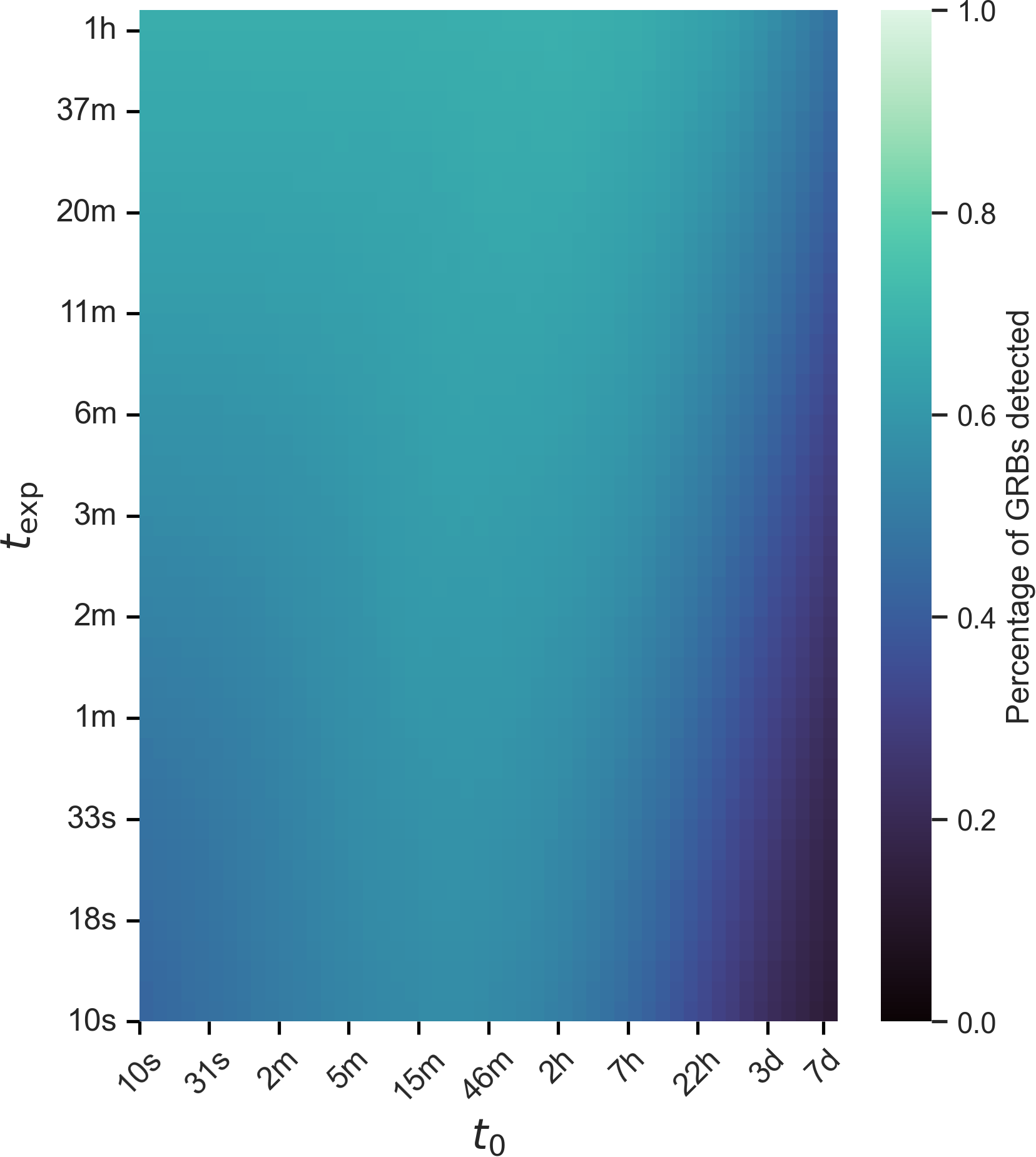}}
  \caption{Overall detectability of short GRBs with CTA South (upper panels) and CTA North (lower panels) for the entire sample of simulated events given latency $t_{0}$ and exposure time $t_{\text{exp}}$. The left panels show a subset of the sources with viewing angle $\theta_{\text{view}} < 10^{\circ}$, while the right panels show all sources with $\theta_{\text{view}} < 45^{\circ}$. The color scale indicates the fraction of simulated sources detected, where lighter regions correspond to higher detection rates.}
  \label{fig:cta_detectability}
\end{figure}

In our simulations, we estimate the exposure time needed for CTA to detect the GRBs as a function of the latency time from the onset of the GRB emission. In all simulations, we assume the so-called CTA Alpha Configuration with a total of 9 telescopes in the north and 51 in the south~\cite{cherenkov_telescope_array_observatory_2021_5499840}. Results using other upgraded configurations will be presented in a future study. The exposure time is calculated as the time required for a $5\sigma$ detection. The minimum fluence detectable by CTA is computed using the Instrument Response Functions (IRFs) from Monte Carlo simulations with the \texttt{ctools} package~\cite{cta_irf_comparisons_2023, 2016ascl.soft01005K}. Given our simulated catalog of BNS merger events, we can estimate the necessary exposure time for a VHE detection based on event parameters such as redshift and $E_{\text{iso}}$, should they become known.


Figure~\ref{fig:cta_detectability} shows the fraction of simulated events detectable by CTA given a latency until the start of observations $t_{0}$ and exposure time $t_{\text{exp}}$. Our simulations show that for on-axis GRBs (viewing angle $< 10^\circ$) and a latency time of about 10 minutes, approximately 98\% of the sources can be detected with an exposure time of one minute. For a shorter latency time of about 30 seconds, nearly all (99\%) of the sources can be detected with an exposure time of one minute or less. When including off-axis GRBs with a viewing angle $< 45^\circ$, about 35\% of the sources can be detected within one minute for a latency time of 10 minutes. These numbers do not yet take into account localization time, moon, or other factors that will affect the true detectability of a given source.

\textbf{Optimizing the CTA observing strategy}

The uncertainties in the source localization for GW events require an efficient observational strategy. In our current approach, we assume the most optimistic scenario where the spectral and temporal evolution of the GRB is known a priori. This allows for optimal observation scheduling for each source. We employ several observation scheduling algorithms to derive optimal pointing patterns that cover the largest total GW uncertainty region possible. These algorithms are part of realistic observation scheduling simulations that take into account visibility conditions, including darkness and moonlight conditions for each GW alert time, as well as prioritization of observations in low zenith angle conditions to achieve lower energy thresholds. The scheduler selects the region with the highest GW source sky-position probability in each iteration of the observation strategy. The exposure is selected based on the exposure time calculation while considering the zenith angle evolution, which is critical for long exposure times, an approach which maximizes the chances of detecting the source.

\begin{figure}[ht]
  \centering
  \subfigure[CTA N+S, On-axis GRBs]{\includegraphics[width=0.45\textwidth]{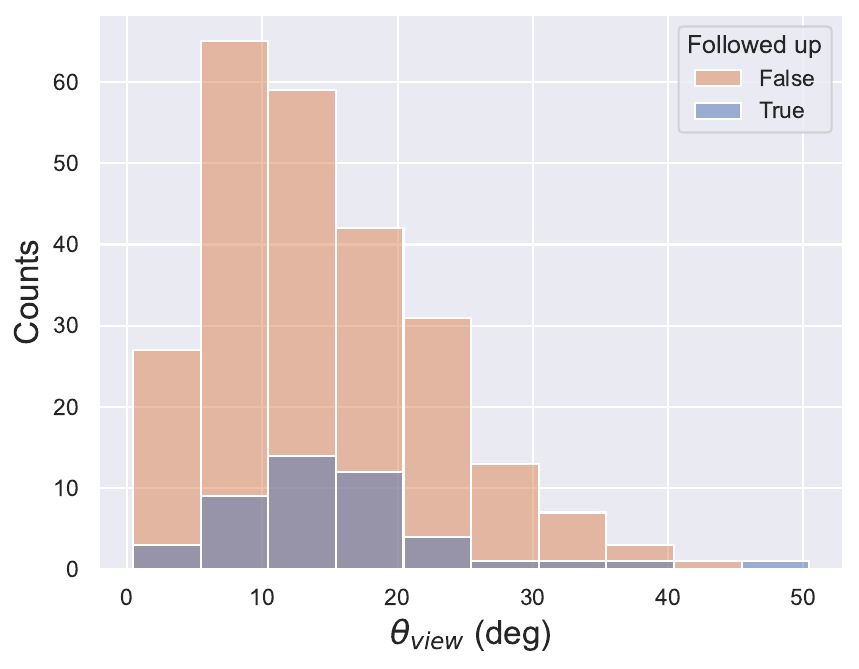}}
  \subfigure[CTA N+S, Off-axis GRBs]{\includegraphics[width=0.45\textwidth]{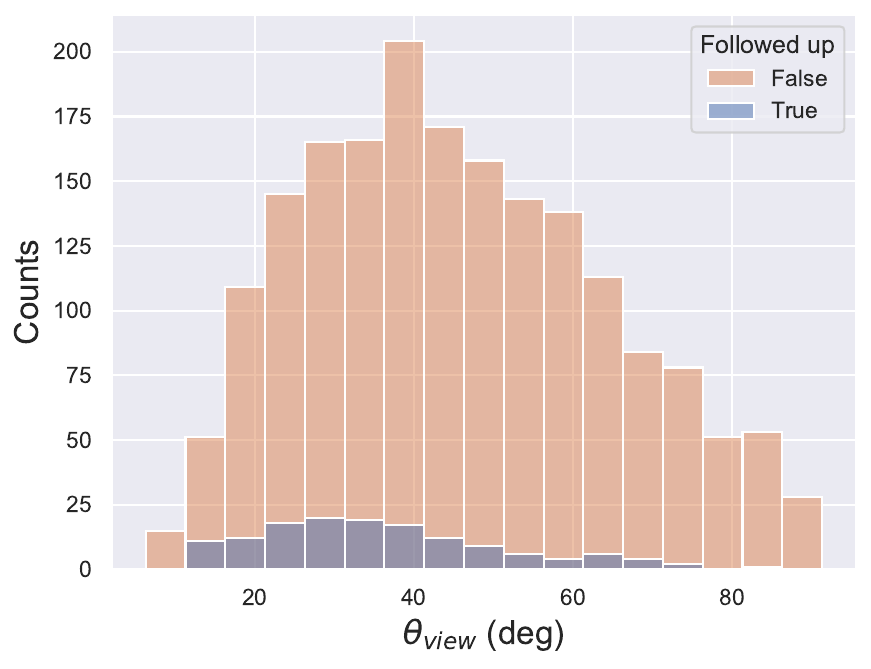}}
  \caption{Number of events for which the observer is located within the light cone of the GRB jet (on-axis, left) and outside of the light cone (off-axis, right) binned by viewing angle $\theta_{\text{view}}$. This is determined by comparing the opening angle of the jet with $\theta_{\text{view}}$ to determine if the observer is located within this opening angle or not. The shorter blue bars show the distribution of events that were followed up by the scheduler, while the taller orange bars show events that were not followed. Events are not scheduled for follow up due to visibility constraints, no region of the skymap being accessible to the telescopes, or if the available time window is not long enough to lead to a $5\sigma$ detection.}
  \label{fig:jet_comparison}
\end{figure}

After running the scheduler on the simulation sets in the catalog, the total number of GW-GRB events that were able to be followed up was ~8\% of the total population and 4.5\% ended up covering the true location of the source. Of the on-axis events (in this case, the observer is within the opening angle of the GRB jet), 18\% were followed up and 10\% covered the true location of the source. For off-axis events, 7\% were followed up and 4\% covered the location of the true source. The distributions of the observable events where the observer is (a) within or (b) outside the lightcone of the jet, on- and off-axis GRBs, respectively, are shown in Figure~\ref{fig:jet_comparison}. Also shown are the sub-populations of these events that are followed up by the scheduler, and therefore possibly detectable by CTA. It can be seen that for on-axis events, those with a viewing angle between $10$ - $20^{\circ}$ are more often followed up, while for off-axis events, this range is in the $20$ - $40^{\circ}$ range.



\section{Discussion}

This work presents an overview of the strategies and simulations involved in the search for the EM counterparts of gravitational waves with CTA. We outline the CTA follow-up strategy, focusing on estimating the exposure times needed for detecting GRBs as a function of the latency time from the onset of the GRB emission. The strategy takes into account the time required to send gravitational wave alerts, the time needed for telescopes to be pointed towards the region of interest, and the uncertainty in the sky location of the gravitational wave event. The scheduler selects the region with the highest gravitational wave source sky-position probability and considers visibility conditions and zenith angle evolution in each iteration of the observation strategy. In addition, the estimated detectability of events with various viewing angles has been simulated. The simulations are promising, especially for the detection of on-axis GRBs, showing that short exposure times at each pointing will suffice in many cases. The computation of the absolute joint GW-CTA detection rate requires the convolution of the fraction of observable events from this catalogue with the assumed BNS merger rate density. This will be the subject of a future work.

The detection of VHE emission associated with gravitational wave events will be a milestone in multimessenger astronomy, providing insights into the mechanisms of short GRBs and the properties of BNS mergers. Through optimized follow-up strategies and continued collaboration with gravitational wave observatories, CTA stands at the forefront of unraveling the mysteries of the most energetic events in the Universe.
\\ \\
\textbf{Acknowledgements:} We gratefully acknowledge financial support from the agencies and organizations listed at \href{https://www.cta-observatory.org/consortium_acknowledgments/}{https://www.cta-observatory.org/consortium\_acknowledgments}. This research has made use of the CTA instrument response functions provided by the CTA Consortium and Observatory, see \href{https://www.cta-observatory.org/science/ctao-performance/}{https://www.cta-observatory.org/science/ctao-performance}  for more details.

%
%
%

\renewcommand*{\bibfont}{\small}
\bibliographystyle{JHEP}
\bibliography{biblio}

\providecommand{\href}[2]{#2}\begingroup\raggedright\begin{thebibliography}{10}

\bibitem{2015CQGra..32g4001L}
{LIGO Scientific Collaboration}, \emph{{Advanced LIGO}},
  \href{https://doi.org/10.1088/0264-9381/32/7/074001}{\emph{Classical and
  Quantum Gravity} {\bfseries 32} (2015) 074001}
  [\href{https://arxiv.org/abs/1411.4547}{{\ttfamily 1411.4547}}].

\bibitem{2015CQGra..32b4001A}
F.~{Acernese}, M.~{Agathos}, K.~{Agatsuma}, D.~{Aisa}, N.~{Allemandou},
  A.~{Allocca} et~al., \emph{{Advanced Virgo: a second-generation
  interferometric gravitational wave detector}},
  \href{https://doi.org/10.1088/0264-9381/32/2/024001}{\emph{Classical and
  Quantum Gravity} {\bfseries 32} (2015) 024001}
  [\href{https://arxiv.org/abs/1408.3978}{{\ttfamily 1408.3978}}].

\bibitem{2017ApJ...848L..14G}
A.~{Goldstein}, P.~{Veres}, E.~{Burns}, M.S.~{Briggs}, R.~{Hamburg},
  D.~{Kocevski} et~al., \emph{{An Ordinary Short Gamma-Ray Burst with
  Extraordinary Implications: Fermi-GBM Detection of GRB 170817A}},
  \href{https://doi.org/10.3847/2041-8213/aa8f41}{\emph{\apjl} {\bfseries 848}
  (2017) L14} [\href{https://arxiv.org/abs/1710.05446}{{\ttfamily
  1710.05446}}].

\bibitem{2017ApJ...848L..15S}
V.~{Savchenko}, C.~{Ferrigno}, E.~{Kuulkers}, A.~{Bazzano}, E.~{Bozzo},
  S.~{Brandt} et~al., \emph{{INTEGRAL Detection of the First Prompt Gamma-Ray
  Signal Coincident with the Gravitational-wave Event GW170817}},
  \href{https://doi.org/10.3847/2041-8213/aa8f94}{\emph{\apjl} {\bfseries 848}
  (2017) L15} [\href{https://arxiv.org/abs/1710.05449}{{\ttfamily
  1710.05449}}].

\bibitem{2017ApJ...848L..13A}
B.P.~{Abbott}, R.~{Abbott}, T.D.~{Abbott}, F.~{Acernese}, K.~{Ackley},
  C.~{Adams} et~al., \emph{{Gravitational Waves and Gamma-Rays from a Binary
  Neutron Star Merger: GW170817 and GRB 170817A}},
  \href{https://doi.org/10.3847/2041-8213/aa920c}{\emph{\apjl} {\bfseries 848}
  (2017) L13} [\href{https://arxiv.org/abs/1710.05834}{{\ttfamily
  1710.05834}}].

\bibitem{2019Natur.575..455M}
{MAGIC Collaboration}, V.A.~{Acciari}, S.~{Ansoldi}, L.A.~{Antonelli},
  A.~{Arbet Engels}, D.~{Baack} et~al., \emph{{Teraelectronvolt emission from
  the {\ensuremath{\gamma}}-ray burst GRB 190114C}},
  \href{https://doi.org/10.1038/s41586-019-1750-x}{\emph{\nat} {\bfseries 575}
  (2019) 455} [\href{https://arxiv.org/abs/2006.07249}{{\ttfamily
  2006.07249}}].

\bibitem{2021ApJ...908...90A}
V.A.~{Acciari}, S.~{Ansoldi}, L.A.~{Antonelli}, A.~{Arbet Engels}, K.~{Asano},
  D.~{Baack} et~al., \emph{{MAGIC Observations of the Nearby Short Gamma-Ray
  Burst GRB 160821B}},
  \href{https://doi.org/10.3847/1538-4357/abd249}{\emph{\apj} {\bfseries 908}
  (2021) 90} [\href{https://arxiv.org/abs/2012.07193}{{\ttfamily 2012.07193}}].

\bibitem{2020GCN.29075....1B}
O.~{Blanch}, F.~{Longo}, A.~{Berti}, S.~{Fukami}, Y.~{Suda}, S.~{Loporchio}
  et~al., \emph{{GRB 201216C: MAGIC detection in very high energy gamma rays}},
  {\emph{GRB Coordinates Network} {\bfseries 29075} (2020) 1}.

\bibitem{2020GCN.28659....1B}
O.~{Blanch}, M.~{Gaug}, K.~{Noda}, A.~{Berti}, E.~{Moretti}, D.~{Miceli}
  et~al., \emph{{MAGIC observations of GRB 201015A: hint of very high energy
  gamma-ray signal}}, {\emph{GRB Coordinates Network} {\bfseries 28659} (2020)
  1}.

\bibitem{2019Natur.575..464A}
H.~{Abdalla}, R.~{Adam}, F.~{Aharonian}, F.~{Ait Benkhali}, E.O.~{Ang{\"u}ner},
  M.~{Arakawa} et~al., \emph{{A very-high-energy component deep in the
  {\ensuremath{\gamma}}-ray burst afterglow}},
  \href{https://doi.org/10.1038/s41586-019-1743-9}{\emph{\nat} {\bfseries 575}
  (2019) 464} [\href{https://arxiv.org/abs/1911.08961}{{\ttfamily
  1911.08961}}].

\bibitem{2021Sci...372.1081H}
{H.~E.~S.~S. Collaboration}, H.~{Abdalla}, F.~{Aharonian}, F.~{Ait Benkhali},
  E.O.~{Ang{\"u}ner}, C.~{Arcaro} et~al., \emph{{Revealing x-ray and gamma ray
  temporal and spectral similarities in the GRB 190829A afterglow}},
  \href{https://doi.org/10.1126/science.abe8560}{\emph{Science} {\bfseries 372}
  (2021) 1081} [\href{https://arxiv.org/abs/2106.02510}{{\ttfamily
  2106.02510}}].

\bibitem{2022GCN.32677....1H}
Y.~{Huang}, S.~{Hu}, S.~{Chen}, M.~{Zha}, C.~{Liu}, Z.~{Yao} et~al.,
  \emph{{LHAASO observed GRB 221009A with more than 5000 VHE photons up to
  around 18 TeV}}, {\emph{GRB Coordinates Network} {\bfseries 32677} (2022) 1}.

\bibitem{2020LRR....23....3A}
B.P.~{Abbott}, R.~{Abbott}, T.D.~{Abbott}, S.~{Abraham}, F.~{Acernese},
  K.~{Ackley} et~al., \emph{{Prospects for observing and localizing
  gravitational-wave transients with Advanced LIGO, Advanced Virgo and KAGRA}},
  \href{https://doi.org/10.1007/s41114-020-00026-9}{\emph{Living Reviews in
  Relativity} {\bfseries 23} (2020) 3}.

\bibitem{2022ApJ...924...54P}
P.~{Petrov}, L.P.~{Singer}, M.W.~{Coughlin}, V.~{Kumar}, M.~{Almualla},
  S.~{Anand} et~al., \emph{{Data-driven Expectations for Electromagnetic
  Counterpart Searches Based on LIGO/Virgo Public Alerts}},
  \href{https://doi.org/10.3847/1538-4357/ac366d}{\emph{\apj} {\bfseries 924}
  (2022) 54} [\href{https://arxiv.org/abs/2108.07277}{{\ttfamily 2108.07277}}].

\bibitem{2022MNRAS.513.4159P}
B.~{Patricelli}, M.G.~{Bernardini}, M.~{Mapelli}, P.~{D'Avanzo},
  F.~{Santoliquido}, G.~{Cella} et~al., \emph{{Prospects for multimessenger
  detection of binary neutron star mergers in the fourth LIGO-Virgo-KAGRA
  observing run}}, \href{https://doi.org/10.1093/mnras/stac1167}{\emph{\mnras}
  {\bfseries 513} (2022) 4159}
  [\href{https://arxiv.org/abs/2204.12504}{{\ttfamily 2204.12504}}].

\bibitem{2021zndo...5206853S}
L.~{Singer}, ``{Data-driven expectations for electromagnetic counterpart
  searches based on LIGO/Virgo public alerts: cluster scripts}.'' Zenodo, Aug.,
  2021.
\newblock 10.5281/zenodo.5206853.

\bibitem{2016ARA&A..54..401O}
F.~{{\"O}zel} and P.~{Freire}, \emph{{Masses, Radii, and the Equation of State
  of Neutron Stars}},
  \href{https://doi.org/10.1146/annurev-astro-081915-023322}{\emph{\araa}
  {\bfseries 54} (2016) 401}
  [\href{https://arxiv.org/abs/1603.02698}{{\ttfamily 1603.02698}}].

\bibitem{2016A&A...594A..13P}
{Planck Collaboration}, P.A.R.~{Ade}, N.~{Aghanim}, M.~{Arnaud}, M.~{Ashdown},
  J.~{Aumont} et~al., \emph{{Planck 2015 results. XIII. Cosmological
  parameters}}, \href{https://doi.org/10.1051/0004-6361/201525830}{\emph{\aap}
  {\bfseries 594} (2016) A13}
  [\href{https://arxiv.org/abs/1502.01589}{{\ttfamily 1502.01589}}].

\bibitem{2016PhRvD..93b4013S}
L.P.~{Singer} and L.R.~{Price}, \emph{{Rapid Bayesian position reconstruction
  for gravitational-wave transients}},
  \href{https://doi.org/10.1103/PhysRevD.93.024013}{\emph{\prd} {\bfseries 93}
  (2016) 024013} [\href{https://arxiv.org/abs/1508.03634}{{\ttfamily
  1508.03634}}].

\bibitem{2016A&A...594A..84G}
G.~{Ghirlanda}, O.S.~{Salafia}, A.~{Pescalli}, G.~{Ghisellini},
  R.~{Salvaterra}, E.~{Chassande-Mottin} et~al., \emph{{Short gamma-ray bursts
  at the dawn of the gravitational wave era}},
  \href{https://doi.org/10.1051/0004-6361/201628993}{\emph{\aap} {\bfseries
  594} (2016) A84} [\href{https://arxiv.org/abs/1607.07875}{{\ttfamily
  1607.07875}}].

\bibitem{2003ApJ...592.1002S}
J.D.~{Salmonson}, \emph{{Perspective on Afterglows: Numerically Computed Views,
  Light Curves, and the Analysis of Homogeneous and Structured Jets with
  Lateral Expansion}}, \href{https://doi.org/10.1086/375580}{\emph{\apj}
  {\bfseries 592} (2003) 1002}
  [\href{https://arxiv.org/abs/astro-ph/0307525}{{\ttfamily
  astro-ph/0307525}}].

\bibitem{2017MNRAS.472.4953L}
G.P.~{Lamb} and S.~{Kobayashi}, \emph{{Electromagnetic counterparts to
  structured jets from gravitational wave detected mergers}},
  \href{https://doi.org/10.1093/mnras/stx2345}{\emph{\mnras} {\bfseries 472}
  (2017) 4953} [\href{https://arxiv.org/abs/1706.03000}{{\ttfamily
  1706.03000}}].

\bibitem{cherenkov_telescope_array_observatory_2021_5499840}
C.T.A.~Observatory and C.T.A.~Consortium, \emph{{CTAO Instrument Response
  Functions - prod5 version v0.1}},
  \href{https://doi.org/10.5281/zenodo.5499840}{\emph{Zenodo} (2021) }.

\bibitem{cta_irf_comparisons_2023}
G.~{Maier}, O.~{Gueta} and R.~{Zanin}, \emph{{CTAO Instrument Response
  Functions: Comparison of prod5 and prod3b releases}},
  \href{https://doi.org/10.5281/zenodo.8050921}{\emph{Zenodo} (2023) }.

\bibitem{2016ascl.soft01005K}
J.~{Kn{\"o}dlseder}, M.~{Mayer}, C.~{Deil}, R.~{Buehler}, J.~{Bregeon} and
  P.~{Martin}, ``{ctools: Cherenkov Telescope Science Analysis Software}.''
  Astrophysics Source Code Library, record ascl:1601.005, Jan., 2016.

\end{thebibliography}\endgroup

\end{document}